\documentclass[%
 aip,
 amsmath,amssymb,
 reprint,%
]{revtex4-1}

\usepackage{graphicx}
\usepackage{dcolumn}
\usepackage{bm}

\usepackage{amsmath}
\usepackage{amssymb}
\usepackage{graphicx}
\usepackage{indentfirst}
\usepackage{booktabs}
\usepackage{multirow}
\usepackage{colortbl}

\usepackage[utf8]{inputenc}
\usepackage[T1]{fontenc}
\usepackage{mathptmx}
\usepackage{etoolbox}

\makeatletter
\def\@email#1#2{%
 \endgroup
 \patchcmd{\titleblock@produce}
  {\frontmatter@RRAPformat}
  {\frontmatter@RRAPformat{\produce@RRAP{*#1\href{mailto:#2}{#2}}}\frontmatter@RRAPformat}
  {}{}
}%
\makeatother
\begin{document}

\preprint{AIP/123-QED}

\title{Valley Polarization and Anomalous Valley Hall Effect in Altermagnet Ti$_2$Se$_2$S with Multipiezo Properties}

\author{Xin Hu}
\thanks{These authors contributed equally to this work.}
\affiliation{State Key Laboratory for Mechanical Behavior of Materials, Center for Spintronics and Quantum System, School of Materials Science and Engineering, Xi'an Jiaotong University, Xi'an, Shaanxi, 710049, China}
\author{Weihang Zhao}
\thanks{These authors contributed equally to this work.}
\affiliation{State Key Laboratory for Mechanical Behavior of Materials, Center for Spintronics and Quantum System, School of Materials Science and Engineering, Xi'an Jiaotong University, Xi'an, Shaanxi, 710049, China}
\author{Wenjun Xia}
\thanks{These authors contributed equally to this work.}
\affiliation{State Key Laboratory for Mechanical Behavior of Materials, Center for Spintronics and Quantum System, School of Materials Science and Engineering, Xi'an Jiaotong University, Xi'an, Shaanxi, 710049, China}
\author{Hanbo Sun}
\affiliation{State Key Laboratory for Mechanical Behavior of Materials, Center for Spintronics and Quantum System, School of Materials Science and Engineering, Xi'an Jiaotong University, Xi'an, Shaanxi, 710049, China}
\author{Chao Wu}
\affiliation{State Key Laboratory for Mechanical Behavior of Materials, Center for Spintronics and Quantum System, School of Materials Science and Engineering, Xi'an Jiaotong University, Xi'an, Shaanxi, 710049, China}
\author{Yin-Zhong Wu}
\email{yzwu@usts.edu.cn}
\altaffiliation[Also at ]{Authors to whom correspondence should be addressed: yzwu@usts.edu.cn}
\affiliation{School of Physical Science and Technology, Suzhou University of Science and Technology, Suzhou 215009, China }
\author{Ping Li}
\email{pli@xjtu.edu.cn}
\altaffiliation[Also at ]{Authors to whom correspondence should be addressed: pli@xjtu.edu.cn}
\affiliation{State Key Laboratory for Mechanical Behavior of Materials, Center for Spintronics and Quantum System, School of Materials Science and Engineering, Xi'an Jiaotong University, Xi'an, Shaanxi, 710049, China}
\affiliation{State Key Laboratory of Silicon and Advanced Semiconductor Materials, Zhejiang University, Hangzhou, 310027, China}
\affiliation{State Key Laboratory for Surface Physics and Department of Physics, Fudan University, Shanghai, 200433, China}

\date{\today}

\begin{abstract}
Recently, altermagnets demonstrate numerous newfangle physical phenomena due to their inherent antiferromagnetic coupling and spontaneous spin splitting, that are anticipated to enable innovative spintronic devices. However, the rare two-dimensional altermagnets have been reported, making it difficult to meet the requirements for high-performance spintronic devices on account of the growth big data. Here, we predict a stable monolayer Ti$_2$Se$_2$S with out-of-plane altermagnetic ground state and giant valley splitting. The electronic properties of altermagnet Ti$_2$Se$_2$S are highly dependent on the onsite electron correlation. Through symmetry analysis, we find that the valleys of X and Y points are protected by the mirror M$_{xy}$ symmetry rather than the time-reversal symmetry. Therefore, the multipiezo effect, including piezovalley and piezomagnetism, can be induced by the uniaxial strain. The total valley splitting of monolayer Ti$_2$Se$_2$S can be as high as $\sim$500 meV. Most interestingly, the direction of valley polarization can be effectively tuned by the uniaxial strain, based on this, we have defined logical "0", "+1", and "-1" states for data transmission and storage. In addition, we have designed a schematic diagram for observing the anomalous Hall effect in experimentally. Our findings have enriched the candidate materials of two-dimensional altermagnet for the ultra-fast and low power consumption device applications.
\end{abstract}

\maketitle

\maketitle

Altermagnetism, a newly discovered magnetic phase, brought about widespread attention due to the spin splitting and zero net magnetization \cite{1,2,3,4}. Importantly, the emergence of these novel phenomena does not require the spin-orbit coupling (SOC) effect \cite{1,2}. In addition, the spin-dependent Fermi surface in altermagnets exhibit the planar or bulk $\emph{d}$-wave, $\emph{g}$-wave, or $\emph{i}$-wave symmetry in momentum space \cite{1,2}. It indicates that the altermagnet has the virtues of resisting external field perturbations,switching speed, ultrafast spin dynamics of antiferromagnet (AFM) and intrinsic ferromagnetic spin splitting. These features of altermagnetic materials can induce a series of novel quantum phenomena. Such as, the spin splitting induced spin current generation \cite{5,6,7}; the large anomalous Hall effect comparable to that of FM \cite{8}; the significant spin Seebeck, crystal Nernst and crystal thermal Hall effects \cite{9,10}; the staggered spin-momentum interaction caused by the time-reversal symmetry breaking \cite{11}; the tunneling and giant magnetoresistance effect \cite{12}; the crystal chirality magneto-optical response \cite{13}; the nonlinear transport \cite{14}; the theoretically proposed spin-splitter torque \cite{15} and experimentally confirmed \cite{16,17}; the nontrivial superconductivity \cite{18,19}; which meaning broad application prospects. To date, the experimental investigation mainly focuses on three-dimensional materials such as MnTe \cite{4,20,21}, RuO$_2$ \cite{8,17}, CrSb \cite{22,23,24}, and Cr-doped FeSb$_2$ \cite{25}. When quantum is confined to a two-dimensional (2D) system, more abundant physical phenomena will emerge \cite{26,27}. Unfortunately, there are few 2D altermagnets suitable for experimental research, which is an urgent need to find more altermagnets.

Valley represents a third independent quantum degree of freedom for electrons, existing alongside charge and spin properties, which has caused extensive concern since it offers exceptional potential for realizing future devices featuring THz-speed operation, unprecedented capacity, ultra-low power consumption, and nonvolatile data retention \cite{28,29,30}. The valley index corresponds to the local energy extremal points in the band structure. To utilize the valley index as an information encoding parameter, controlled manipulation of valley carriers is essential to achieve valley polarization and realize anomalous valley Hall effect. Currently, there are two methods to achieve spontaneous valley polarization. One method is to break inversion symmetry ($\emph{P}$) through ferroelectricity \cite{30}, while the other is to break time-reversal symmetry ($\emph{T}$) by magnetism \cite{31,32,33,34,35,36}. They are named ferroelectric ferrovalley and magnetic ferrovalley. However, the altermagnets that also broken the $\emph{T}$ symmetry, how does it realize valley polarization?

In this work, based on the first-principles calculations, we predict that the 2D altermagnet Ti$_2$Se$_2$S is a candidate material with promising application prospects in multipiezo and valleytronic. Monolayer Ti$_2$Se$_2$S exhibits stable out-of-plane altermagent properties. It shows characteristics of semiconductor with the band gap located at the X and Y points. Interestingly, the valence band maximum (VBM) and conduction band minimum (CBM) at the X point are spin down and spin up bands, respectively, while it is just the opposite at Y point. Moreover, the transformation from metal to semiconductor is demonstrated with the increase of then Hubbard U value. Our results demonstrate that the uniaxial strain can effectively tune the magnitude and direction of valley polarization. In addition, the abundant multipiezo effect, including the piezoelectric and piezomagnetism, can be realized in monolayer Ti$_2$Se$_2$S. Based on these, we designed the devices of anomalous valley Hall effect and piezoelectric effect. The unique combination of physical properties in monolayer Ti$_2$Se$_2$S makes it a highly promising candidate material for multifunctional valleytronic and spintronic device applications.

Within the framework of density functional theory (DFT), we systematically explored the magnetic and  electronic properties through the Vienna ab initio Simulation Package (VASP) \cite{37,38,39}. The exchange-correlation energy was treated within the Perdew-Burke-Ernzerhof (PBE) of the generalized gradient approximation (GGA) \cite{40}. The $21\times 21\times 1$ $\Gamma$-centered $k$ meshes of Brillouin zone (BZ) is adopted. The plane-wave basis set with a kinetic energy cutoff of 500 eV is employed. The structural optimizations is performed with convergence criterion of 10$^{-6}$ eV for total energy and -0.01 eV/{\AA} for Hellmann-Feynman forces. A 20 {\AA} vacuum layer is added perpendicular to the 2D plane (c-axis direction) in the slab geometry, effectively suppressing spurious interactions between the monolayer and its periodic replicas. To describe strongly correlated 3d electrons of Ti \cite{41}, the GGA + U method is used with the Coulomb repulsion U value of 3.0 $\sim$ 4.0 eV.

Monolayer Ti$_2$Se$_2$S behaves a 2D square lattice with the point group of D$_{4h}$ and the space group of P4/mmm, as shown in Fig. 1(a). The crystal structure consists of three atomic layers, which the Ti-S atomic plane are sandwiched between two Se atomic planes, similar to the monolayer V$_2$Se$_2$O \cite{6,42}. The yellow dotted square indicates the unit cell. It shows a M$_{xy}$ mirror and C$_4$ rotational symmetries, that is an important condition required for altermagnet. The Ti atom bonds with the surrounding four Se atoms and two S atoms, forming an octahedral crystal field. The lattice constant of monolayer Ti$_2$Se$_2$S is 4.53 {\AA}, while the bond length of Ti-S and Ti-Se is 2.26 {\AA} and 2.79 {\AA}, respectively. As shown in Fig. 1(c), it exhibits the first Brillouin zone (BZ) including the high symmetry points. It is worth noting that the bond angle of Ti-S-Ti is exactly 180 $^ \circ$, indicating a favorable AFM coupling according to the Goodenough-Kanamori-Anderson rule \cite{43,44,45}. We evaluated the stability of monolayer Ti$_2$Se$_2$S from two aspects. On the one hand, we estimated its thermodynamic stability by $\emph{ab initio}$ molecular dynamics (AIMD). As shown in Fig. 1(b), the total energy of monolayer Ti$_2$Se$_2$S shows minimal fluctuation during 5 ps at 300 K, demonstrating its excellent thermal stability. On the other hand, we calculated the phonon spectrum using $3\times 3\times 1$ supercells to evaluate the dynamic stability. Fig. 1(d) shows the absence of imaginary frequencies, indicating that the monolayer Ti$_2$Se$_2$S is dynamical stability.

\begin{figure}[htb]
\begin{center}
\includegraphics[angle=0,width=1.0\linewidth]{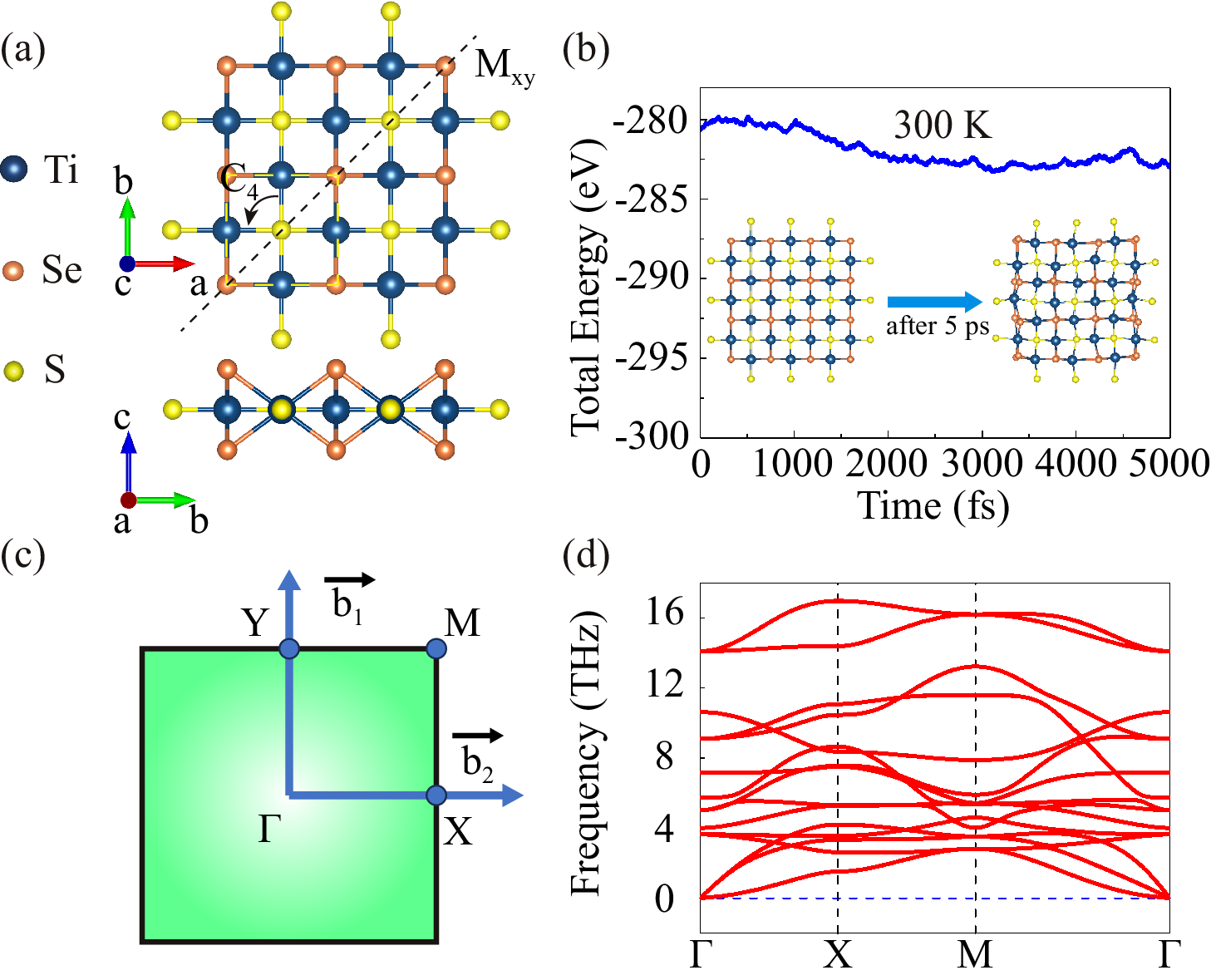}
\caption{(a) The top and side views of monolayer Ti$_2$Se$_2$S structure. The yellow dotted square denotes the unit cell. The cyan, orange, and yellow balls represent Ti, Se, and S elements, respectively. (b) The total energy fluctuation of monolayer Ti$_2$Se$_2$S during 5 ps AIMD simulation at 300 K. The final structures after AIMD simulation are exhibited in the insets. (c) The BZ of the square lattice, characterized by its reciprocal lattice vectors $\vec{b}_1$ and $\vec{b}_2$. The $\Gamma$, X, Y, and M are high-symmetry points in the BZ.  (d) The phonon dispersion curves were calculated along the high-symmetry lines of the BZ.
}
\end{center}
\end{figure}

To confirm the magnetic ground state of monolayer Ti$_2$Se$_2$S, we considered two typical magnetic configurations, namely the AFM and ferromagnetic (FM) states. The calculation results show that the total energy of the AFM state is 66.53 meV lower than that of the FM state, indicating that the AFM is magnetic ground state. Fig. 2(a) shows its magnetic ground state configuration structure. The important factor for the stable existence of 2D magnetic materials is the presence of out-of-plane magnetic anisotropy energy (MAE). The out-of-plane MAE can effectively suppress the magnetic moment fluctuations caused by thermal disturbances and maintain the long-range magnetic order. The MAE is primarily derived from SOC interactions \cite{46}. The MAE is defined as MAE = E$_{100}$ - E$_{001}$, where E$_{100}$ and E$_{001}$ denotes the total energy of the Ti atoms magnetic moment along [100] and [001] directions, respectively. The MAE of monolayer Ti$_2$Se$_2$S is 0.77 meV, which indicates that the direction of easy magnetization is along the out-of-plane. Besides, for the octahedral symmetry of monolayer Ti$_2$Se$_2$S, the MAE can be written in the form of angle dependence,
\begin{equation}
\rm MAE= K_1 cos^2\theta + K_2 cos^4\theta,
\end{equation}
where K$_1$, K$_2$, and $\theta$ are the anisotropy constants and azimuthal angle of rotation, respectively. If K$_1$ $<$ 0, it indicates that the easy magnetization direction along the out-of-plane (z-axis), while K$_1$ $>$ 0 suggests that it benefits to be parallel the in-plane (x-axis). The MAE of monolayer Ti$_2$Se$_2$S shows a good fit of Eq (1) as presented in Fig. 2(b), it suggests a pronounced dependence of the MAE on the magnetization orientation within the xz plane.

\begin{figure}[htb]
\begin{center}
\includegraphics[angle=0,width=1.0\linewidth]{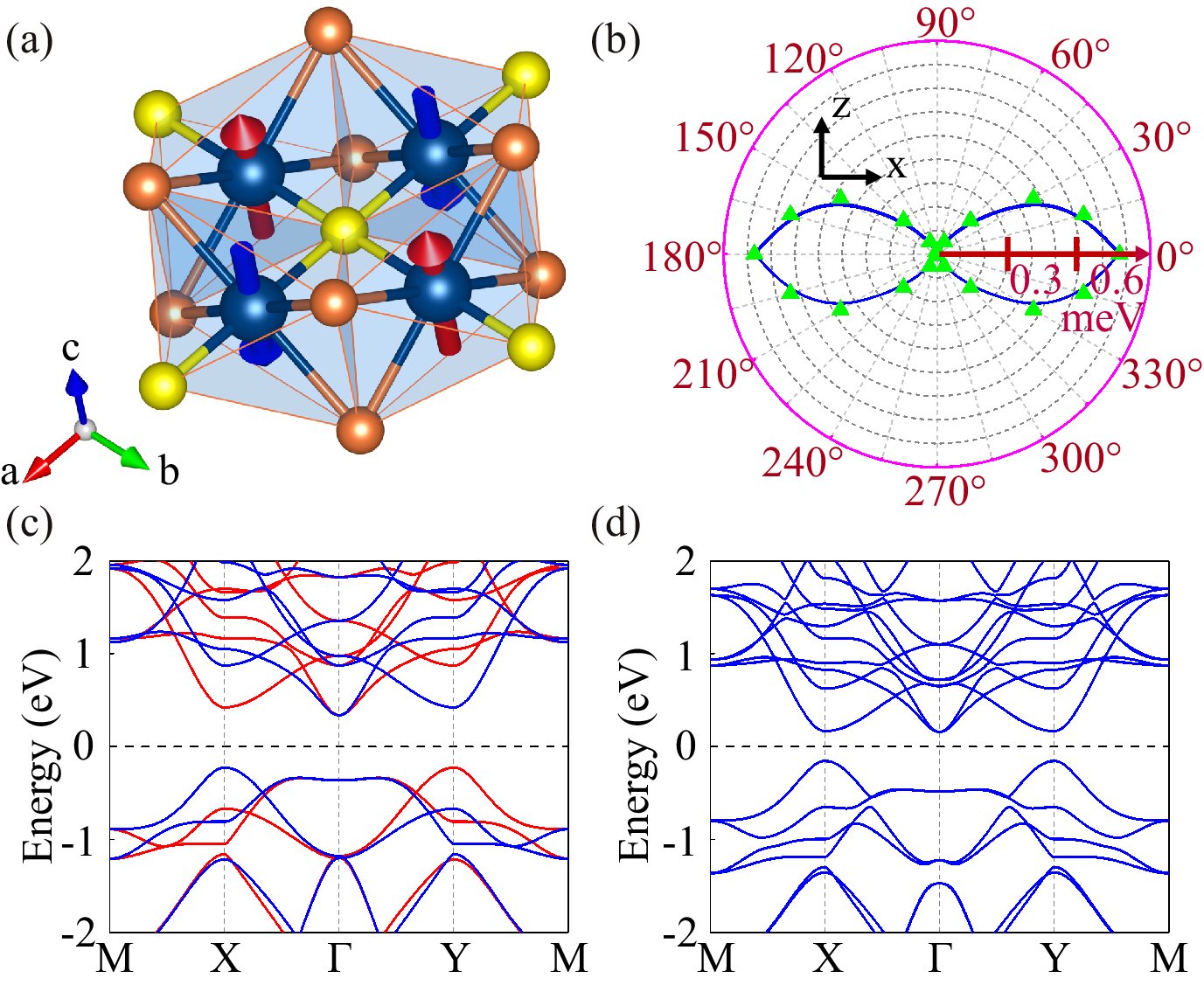}
\caption{(a) The magnetic configuration of monolayer Ti$_2$Se$_2$S and the octahedral crystal field. (b) The angular variation of MAE in monolayer Ti$_2$Se$_2$S, when the magnetization orientation is confined within the xz plane. (c) Spin-polarized band structure of monolayer Ti$_2$Se$_2$S. The red and blue lines denote spin up and spin down bands, respectively. (d) Band structure of monolayer Ti$_2$Se$_2$S with the SOC effect.
}
\end{center}
\end{figure}

For the traditional AFM system with the combined of $\emph{T}$ and $\emph{P}$ symmetry ($\emph{PT}$), which connects the energy eigenvalues E$_{\uparrow}$($\textbf{k}$) and E$_{\downarrow}$($\textbf{k}$), ensuring the spin degeneracy. The $\emph{P}$ operation only converse the vector $\emph{k}$. Therefore, it can be ensured that the eigenvalues are satisfied $\emph{P}$E$_{\uparrow}$($\emph{k}$) = E$_{\uparrow}$($\emph{-k}$). However, the $\emph{T}$ operation can reverse not only $\emph{k}$ but also spin, leading to the $\emph{T}$E$_{\uparrow}$($\emph{k}$) = E$_{\downarrow}$($\emph{-k}$). It indicates that the $\emph{PT}$ symmetry make certain E$_{\uparrow}$($\emph{k}$) = $\emph{PT}$E$_{\uparrow}$($\emph{k}$) = E$_{\downarrow}$($\emph{k}$). The result presented is spin degenerate bands for the two opposite components. In addition, the translation operation ($\emph{t}$) also produces $\emph{t}$E$_{\uparrow}$($\emph{k}$) = E$_{\uparrow}$($\emph{k}$). Therefore, the energy eigenvalue of traditional AFM system satisfies $\emph{PTt}$E$_{\uparrow}$($\emph{k}$) = E$_{\uparrow}$($\emph{k}$). Noted that the spin and real spaces is completely decoupled in the ignoring SOC effect. Consequently, the spin reversal operation $\emph{U}$ generate $\emph{U}$E$_{\uparrow}$($\emph{k}$) = E$_{\uparrow}$($\emph{k}$), where $\emph{U}$ is exclusively defined for collinear spin configurations.

Through the above analysis, the spin splitting of monolayer altermagnet originates from the breaking $\emph{Ut}$ and $\emph{PTt}$ symmetry. As shown in Fig. 2(c), the monolayer Ti$_2$Se$_2$S exhibits a semiconductor state with the Hubbard U = 4.0 eV under the absence SOC effect. The valleys of both valence and conduction bands are degenerate at the X and Y points. Moreover, the VBM at the X point corresponds to spin down band, while the CBM is spin up. Conversely, this spin band is inverted at the Y point. When the SOC is included, as shown in Fig. 2(d), these valleys still remain degenerative. Since the valley degeneracy is protected by M$_{xy}$ symmetry rather than T symmetry in the magnetic-ferrovalley materials \cite{47,48,49}. It means that the valley polarization can be only obtained by uniaxial strain, which induces symmetry breaking in the lattice.

\begin{figure}[htb]
\begin{center}
\includegraphics[angle=0,width=1.0\linewidth]{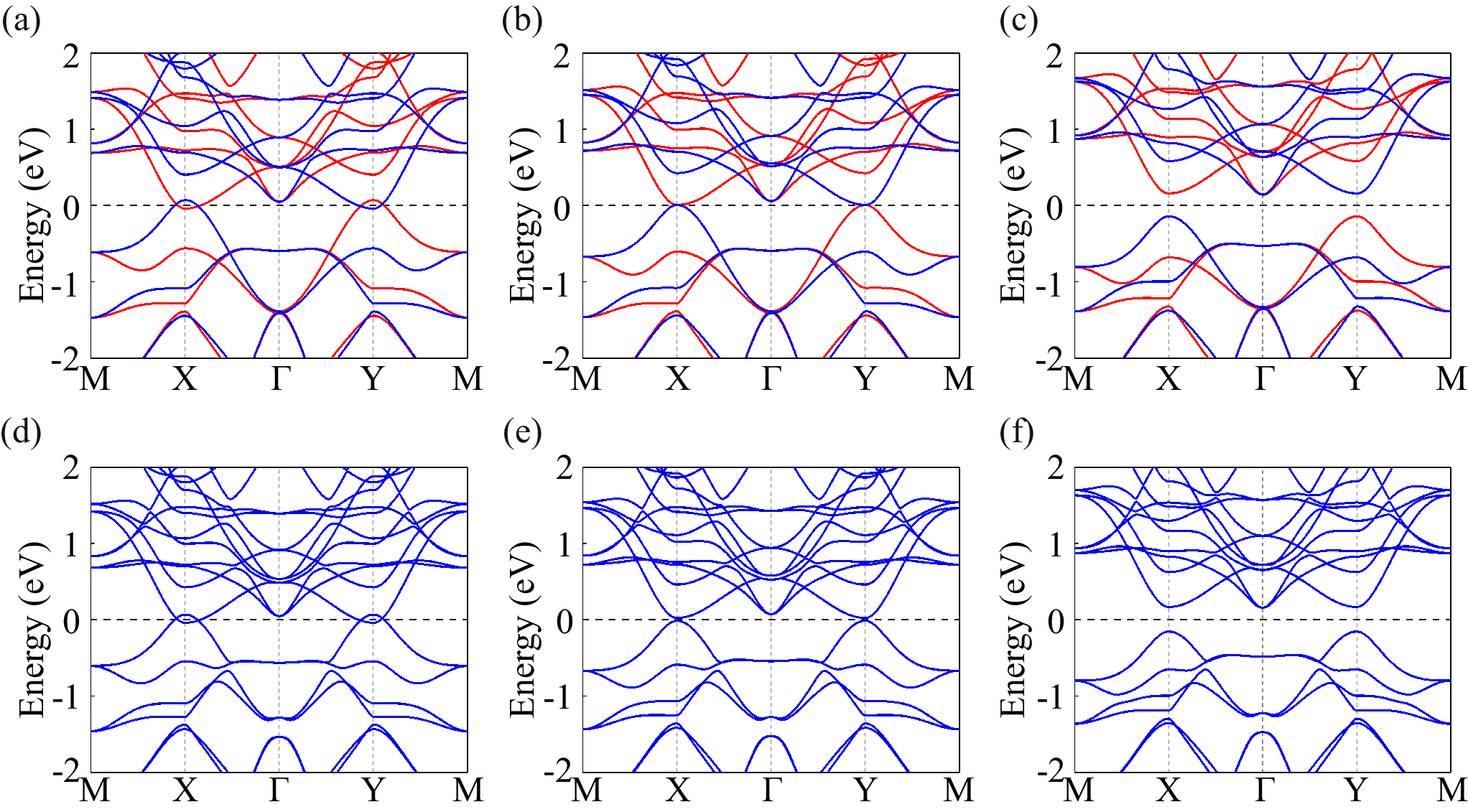}
\caption{(a-c) Spin-polarized band structures of monolayer Ti$_2$Se$_2$S at the different U values, (a) U = 3.0 eV, (b) U = 3.1 eV, and (c) U = 3.5 eV, respectively. The red and blue lines denote spin up and spin down bands, respectively. (d-f) Band structure of monolayer Ti$_2$Se$_2$S with the SOC effect at the different U values, (d) U = 3.0 eV, (e) U = 3.1 eV, and (f) U = 3.5 eV, respectively.
}
\end{center}
\end{figure}

It is well known that the on-site correlation Hubbard U value affects the electronic properties of the system. Therefore, we calculate the band structures of Hubbard U in the range of 3.0 $\sim$ 4.0 eV. Fig. 3(a-c) and Fig. S1 shows the band structures without the SOC effect. When the Hubbard U value is less than 3.1 eV, the monolayer Ti$_2$Se$_2$S is metallic state. The U = 3.1 eV is critical point. As shown in Fig. 3(b), the VBM and CBM just come into contact at the Fermi level. Continuing increasing the Hubbard U value, the X and Y points opens a gap and becomes the semiconductor state. When the SOC is switched on, as shown in Fig. 3(d-f) and Fig. S2, the trend of change is completely consistent without the SOC effect. It is worth noting that the band gap is opened in E$_F$-1.5 eV at $\Gamma$ point, which may have topological properties.

The valleys of monolayer Ti$_2$Se$_2$S are protected by mirror M$_{xy}$ symmetry rather than $\emph{T}$ symmetry. Moreover, the valley polarization dose not depend on the SOC effect. If one wants to realize valley polarization in altermagnetic monolayer Ti$_2$Se$_2$S, it is necessary to break the M$_{xy}$ symmetry through uniaxial strain. Simultaneously, the rotational symmetry of monolayer Ti$_2$Se$_2$S will be lowered from C$_4$ to C$_2$. This valley polarization is named piezovalley. Here, the valley splitting of monolayer Ti$_2$Se$_2$S is defined as the energy difference $\Delta V$ and $\Delta C$ between X and Y at the VBM and CBM, $\Delta C(V)$ = E$_{Xc(v)}$ - E$_{Yc(v)}$. Therefore, as shown in Fig. 4(a-c) and Fig. S3, we investigate the band structure under uniaxial strain along a direction ranging from -5$\%$ $\sim$ 5$\%$ without the SOC effect. When the uniaxial compressive strain is applied, the CBM band rises at X point and the VBM band drops at Y point. Consequently, the significant valley polarization produce in the VBM and CBM. The magnitude of the valley polarization is linearly related to the uniaxial strain. Continuously increased to -5$\%$ uniaxial compressive strain, the monolayer Ti$_2$Se$_2$S turned into a metallic state. While the uniaxial tensile strain is used, the VBM and CBM bands at Y points will be higher than that at X point. When the SOC is included, as shown in Fig. 4(d-f) and Fig. S4, the band structure is not much difference. It is worth noting that the SOC effect has a very minor influence on the valley polarization. Here, we will conduct a comparison by taking -4$\%$ uniaxial compressive strain as an example. In the absence SOC effect, the valley splittings of valence and conduction bands are 315.06 meV and 175.58 meV, respectively. When considering the effect of SOC, the valley splittings of valence and conduction bands become 305.28 meV and 171.68 meV, respectively. Compared with the very small effect of valley splitting caused by the lattice symmetry breaking.

\begin{figure}[htb]
\begin{center}
\includegraphics[angle=0,width=1.0\linewidth]{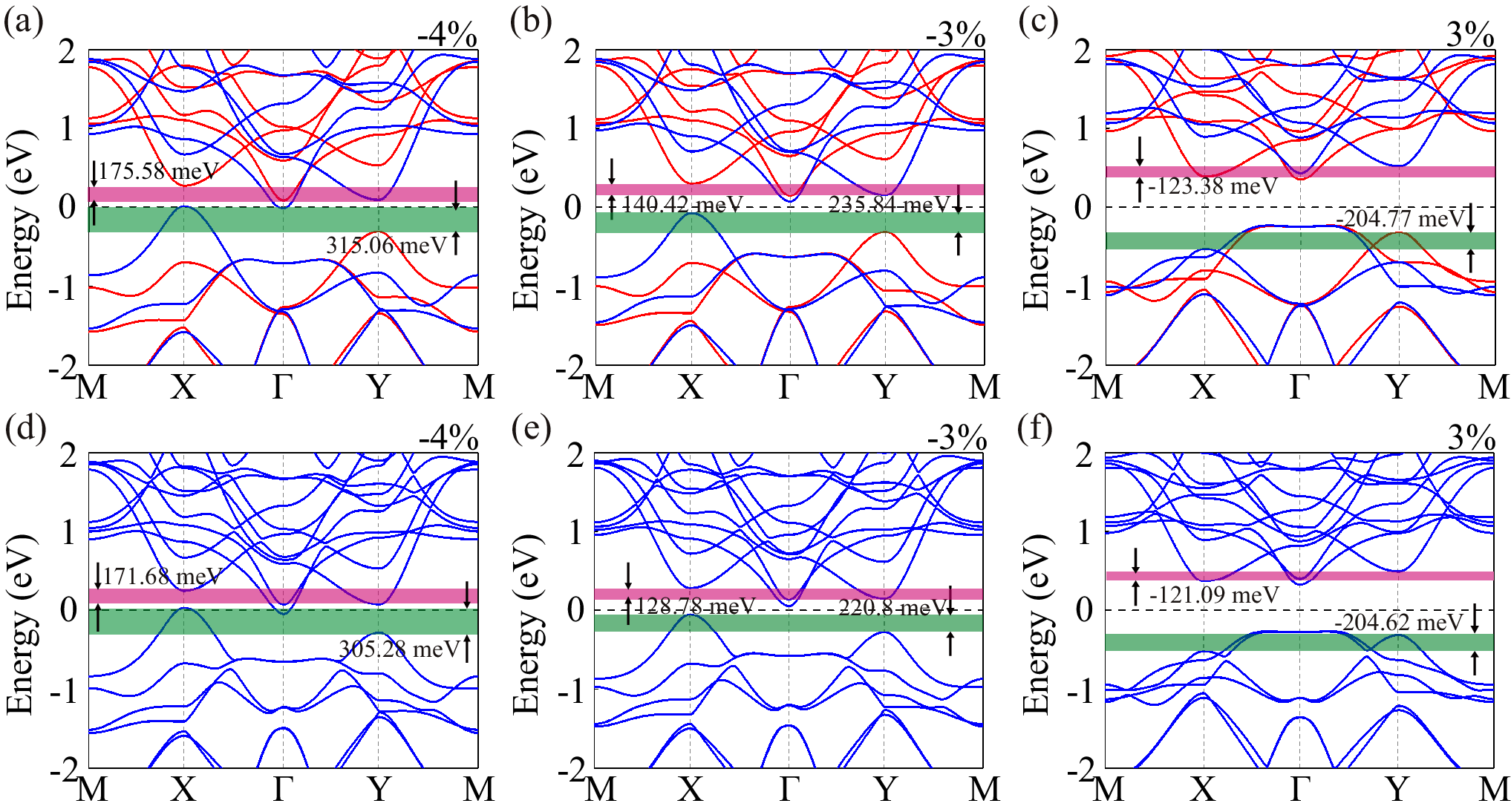}
\caption{ (a-c) Spin-polarized band structures of monolayer Ti$_2$Se$_2$S with the uniaxial strain, (a) -4$\%$, (b) -3$\%$, and (c) 3$\%$, respectively. The red and blue lines denote spin up and spin down bands, respectively. (d-f) Band structure of monolayer Ti$_2$Se$_2$S under the SOC effect with the uniaxial strain, (d) -4$\%$, (e) -3$\%$, and (f) 3$\%$, respectively. The valley splitting of valence and conduction bands are exhibited by green and carmine shading, respectively.
}
\end{center}
\end{figure}

In addition, Fig. 5(a) lists the global band gap variations without and with the SOC effect under the U values ranging from 3.0 $\sim$ 4.0 eV. It further indicates that the trend of band structures changes is consistent without and with the SOC effect. The sole difference lies in U = 3.1 eV. The VBM and CBM are in contact with each other at the Fermi level, exhibiting a metallic state without SOC effect, while it opens the band gap of 38.30 meV with SOC effect. Besides, Fig. 5(b) exhibits the valley splitting under the uniaxial strain along a direction. It is clearly demonstrated that the valley splitting can be significantly tuned under uniaxial strain. The adjustable ranges of valence and conduction bands valley splittings are as high as $\sim$300 meV and $\sim$600 meV, respectively. It is far greater than that of monolayer Fe$_2$Se$_2$O $\sim$350 meV (valence band) \cite{7}, monolayer V$_2$SeTeO $\sim$300 meV (valence band) and $\sim$10 meV (conduction band) \cite{50}, monolayer Nb$_2$SeTeO $\sim$50 meV (valence band) and $\sim$350 meV (conduction band) \cite{51}, and so on.

\begin{figure}[htb]
\begin{center}
\includegraphics[angle=0,width=1.0\linewidth]{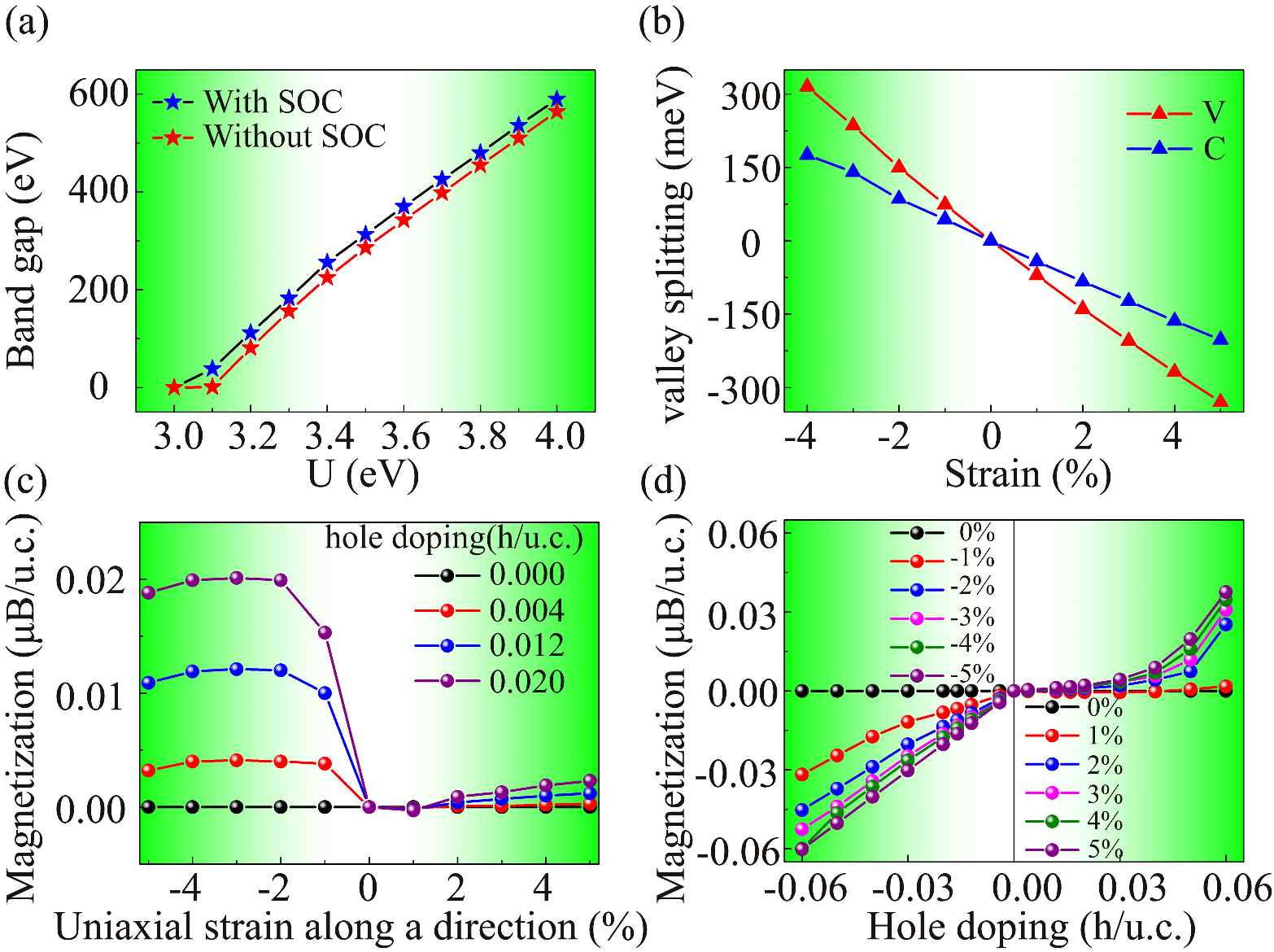}
\caption{ (a) The global band gap with and without the SOC effect at the different U values. (b) Valley splitting of monolayer Ti$_2$Se$_2$S as a function of uniaxial strain along a direction. The valley splitting of valence and conduction bands are signed the V and C, respectively. (c) The net magnetization per unit cell with the various hole doping amounts under the uniaxial strain along a direction. (d) The net magnetization per unit cell with the various hole doping amounts under the certain uniaxial strain along a direction.
}
\end{center}
\end{figure}

In a magnetic material, the system's magnetism is determined by integrating the spin density over all energies up to the Fermi level. Hence, the net magnetic moment can be induced in the monolayer Ti$_2$Se$_2$S by hole or electron doping to shift the Fermi level, such that it crosses only one valley. Based on the symmetry analysis in the above text, we know that the valley splitting of monolayer Ti$_2$Se$_2$S requires a uniaxial strain to break the lattice symmetry. It indicates that the strain-induced valley splitting in the monolayer Ti$_2$Se$_2$S afford the new method to produce net magnetization. Since the magnetization depends on strain, it is named piezomagnetic property. Currently, the piezomagnetism has been only reported in rare 2D materials, such as V$_2$Se$_2$O \cite{6}, Fe$_2$Se$_2$O \cite{7}, and V$_2$SeTeO \cite{42,50}. The net magnetic moment is defined as M = $\int ^{E_f(n)}_{-\infty}[\rho^{\uparrow}(\varepsilon) - \rho^{\downarrow}(\varepsilon)]$dE, where E$_f$, n, $\varepsilon$, and $\rho^{\uparrow(\downarrow)}$ are the doped Fermi level, the doping density, external strain, and spin up (spin down) part of the density of states, respectively. As shown in Fig. 5(c,d), the uniaxial strained monolayer Ti$_2$Se$_2$S appears zero net magnetization without doping, since the number of electrons in the occupied states has not changed. The net magnetization exhibits a linear response under the small strain, reaching a saturation value as strain increases. When doped to a certain concentration, the system exhibits strain-dependent magnetization, increasing under uniaxial tension or compressive strains. It is worth noting that the net magnetization of the previously reported V$_2$Se$_2$O \cite{6}, Fe$_2$Se$_2$O \cite{7}, and V$_2$SeTeO \cite{42,50} is opposite under the uniaxial compressive and tensile strains. The net magnetization of monolayer Ti$_2$Se$_2$S is positive under both uniaxial compressive strain and tensile strain, and the magnetization induced by compressive strain is much larger than that the tensile strain.

According to the above calculations and definitions, the valley polarization are zero, positive, and negative value under without strain, uniaxial compressive strain, and uniaxial tensile strain, respectively. Therefore, based on the valley polarization, as shown in Fig. 6(a-c), we defined the logical "0", "+1", and "-1" states. This implies that the uniaxial strain tune the valley polarization of monolayer Ti$_2$Se$_2$S, it can be used for signal transmission and storage encoding. In addition, the anomalous valley Hall effect of monolayer Ti$_2$Se$_2$S is shown in Fig. 6(d,e). When the uniaxial compressive strain is applied to monolayer Ti$_2$Se$_2$S, as shown in Fig. 6(d). As the Fermi level is tuned to cross between the X and Y points in the VBM band, the spin down holes of X valley will be produced and accumulated on the left edge. When the Fermi level is shifted between the Y and X points in the CBM band, the spin down electrons of Y valley will be generated and accumulated on the right edge. When the compressive strain turns into tensile strain, the situation will be completely reversed. As shown in Fig. 6(e), when the Fermi level once again is moved between the X and Y points in the VBM, the spin up holes of Y valley rather than the spin down holes of X valley will be generated and accumulated on the right edge. Since the uniaxial tensile strain has switched the direction of valley polarization. As the Fermi level is further moved to the region between the X and Y points, the spin up electrons of X valley will be produced and accumulated on the ledge edge. The giant piezovalley effect is anticipated to facilitate multifunction piezovalley application in monolayer Ti$_2$Se$_2$S.

\begin{figure}[htb]
\begin{center}
\includegraphics[angle=0,width=1.0\linewidth]{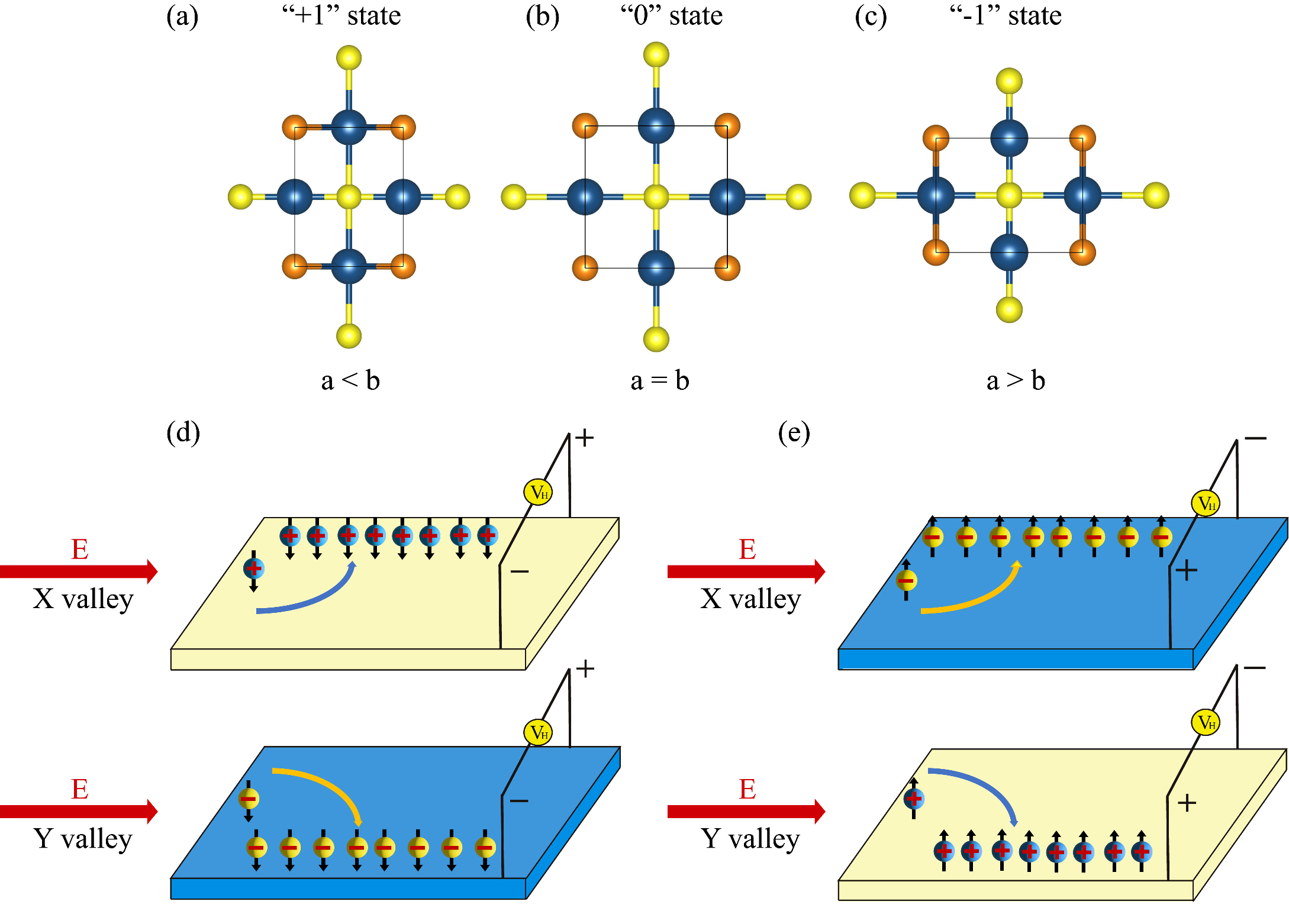}
\caption{ (a-c) Top view of the monolayer Ti$_2$Se$_2$S crystal structure, (a) uniaxial compressive strain along the a direction, (b) without uniaxial strain, (c) uniaxial tensile strain along the a direction. (d, e) Schematic diagram of anomalous valley Hall effect under (d) the uniaxial compressive strain and (e) uniaxial tensile strain at the X and Y valleys. The holes and electrons are represented by the + and - symbol. Spin up carriers are represented by upward arrows, while spin down carriers are indicated by downward arrows.
}
\end{center}
\end{figure}

In conclusion, based on the DFT calculations, we predict that monolayer Ti$_2$Se$_2$S has out-of-plane altermagnetic ground state and giant valley splitting. The band gap of altermagnet monolayer Ti$_2$Se$_2$S are highly dependent on the Hubbard U value. The monolayer Ti$_2$Se$_2$S exhibits metallic state in the Hubbard U < 3.1 eV, while it become a semiconductor state in the Hubbard U > 3.1 eV. The U = 3.1 eV is critical point, the VBM and CBM bands are in contact at the X and Y points of the Fermi level. By symmetry analysis, we find that the X and Y valleys of altermagnet monolayer Ti$_2$Se$_2$S are protected by the mirror M$_{xy}$ symmetry rather than the $\emph{T}$ symmetry. Therefore, the valley polarization and its switching are achieved via the uniaxial strain along a direction, uncovering a piezovalley phenomenon. The total valley splitting of monolayer Ti$_2$Se$_2$S can be as high as $\sim$500 meV. In addition, the net magnetization can be induced in the monolayer Ti$_2$Se$_2$S doped with holes at the uniaxial strain, showing a piezomagnetic effect. Accordingly to the valley polarization, we defined logical "0", "+1", and "-1" states for data transmission and storage, and designed an experimental schematic to detect the anomalous Hall effect. Our work has confirmed that monolayer Ti$_2$Se$_2$S is an ideal multipiezo effect candidate material, the strain engineering offers significant benefits for next-generation multifunctional nanospintronic devices.

~\\
\indent See the supplementary material for the additional results.

~\\
\indent This work is supported by the National Natural Science Foundation of China (Grants No. 12474238, and No. 12004295). P. Li also acknowledge supports from the China's Postdoctoral Science Foundation funded project (Grant No. 2022M722547), and the Open Project of State Key Laboratory of Silicon and Advanced Semiconductor Materials (No. SKL2024-10), and the Open Project of State Key Laboratory of Surface Physics (No. KF2024$\_$02). X. Hu, W. Zhao, and W. Xia thank the National Training Program of Innovation and Entrepreneurship for Undergraduates (Grant No. MSE202410698003).

\section*{AUTHOR DECLARATIONS}
$\textbf{Conflict of Interest}$
The authors have no conflicts to disclose.

$\textbf{Author Contributions}$
Wenjun Xia, Weihang Zhao, and Xin Hu contributed equally to this work.
$\textbf{Xin Hu:}$ Data curation (equal); Formal analysis (equal); Investigation (equal). $\textbf{Weihang Zhao:}$ Data curation (equal); Formal analysis (equal); Investigation (equal); Writing-original draft (supporting). $\textbf{Wenjun Xia:}$ Data curation (equal); Formal analysis (equal); Investigation (equal). $\textbf{Hanbo Sun:}$ Investigation (supporting). $\textbf{Chao Wu:}$ Investigation (supporting). $\textbf{Yin-Zhong Wu:}$ Resources (lead). $\textbf{Ping Li:}$ Conceptualization (lead); Data curation (equal); Formal analysis (equal); Funding acquisition (lead); Investigation (equal); Methodology (lead); Resources (lead); Software (lead); Supervision (equal); Writing-original draft (lead); Writing-review $\&$ editing (lead).

\section*{DATA AVAILABILITY}
The data that support the findings of this study are available from the corresponding authors upon reasonable request.


\nocite{*}
\bibliography{aipsamp}

\section*{REFERENCES}
\begin{thebibliography}{99}


\bibitem{1} L. Smejkal, J. Sinova, and T. Jungwirth, Beyond conventional ferromagnetism and antiferromagnetism: A phase with nonrelativistic spin and crystal rotation symmetry, Phys. Rev. X 12, 031042 (2022).

\bibitem{2} L. Smejkal, J. Sinova, and T. Jungwirth, Emerging Research Landscape of Altermagnetism, Phys. Rev. X 12, 040501 (2022).

\bibitem{3} Y. P. Zhu, X. Chen, X. R. Liu, Y. Liu, P. Liu, H. Zha, G. Qu, C. Hong, J. Li, Z. Jiang, X. M. Ma, Y. J. Hao, M. Y. Zhu, W. Liu, M. Zeng, S. Jayaram, M. Lenger, J. Ding, S. Mo, K. Tanaka, M. Arita, Z. Liu, M. Ye, D. Shen, J. Wrachtrup, Y. Huang, R. H. He, S. Qiao, Q. Liu, and C. Liu, Observation of plaid-like spin splitting in a noncoplanar antiferromagnet, Nature, 626, 523 (2024).

\bibitem{4} J. Krempasky, L. Smejkal, S. W. DSouza, M. Hajlaoui, G. Springholz, K. Uhlirova, F. Alarab, P. C. Constantinou, V. Strocov, D. Usanov, W. R. Pudelko, R. G. Hernandez, A. B. Hellenes, Z. Jansa, H. Reichlova, Z. Soban, R. D. G. Betancourt, P. Wadley, J. Sinova, D. Kriegner, J. Minar, J. H. Dil, and T. Jungwirth, Altermagnetic lifting of Kramers spin degeneracy, Nature, 626, 517 (2024).

\bibitem{5} M. Naka, S. Hayami, H. Kusunose, Y. Yanagi, Y. Motome, and H. Seo, Spin current generation in organic antiferromagnets, Nat. Commun. 10, 4305 (2019).

\bibitem{6} H. Y. Ma, M. Hu, N. Li, J. Liu, W. Yao, J. F. Jia, and J. Liu, Multifunctional antiferromagnetic materials with giant piezomagnetism and noncollinear spin current, Nat. Commun. 12, 2846 (2021).

\bibitem{7} Y. Wu, L. Deng, X. Yin, J. Tong, F. Tian, and X. Zhang, Valley-Related Multipiezo Effect and Noncollinear Spin Current in an Altermagnet Fe$_2$Se$_2$O Monolayer, Nano Lett. 24, 10534 (2024).

\bibitem{8} Z. Feng, X. Zhou, L. Smejkal, L. Wu, Z. Zhu, H. Guo, R. G. Hernandez, X. Wang, H. Yan, P. Qin, X. Zhang, H. Wu, H. Chen, Z. Meng, L. Liu, Z. Xia, J. Sinova, T. Jungwirth, and Z. Liu, An anomalous Hall effect in altermagnetic ruthenium dioxide, Nat. Electron. 5, 735 (2022).

\bibitem{9} Q. Cui, B. Zeng, P. Cui, T. Yu, and H. Yang, Efficient spin Seebeck and spin Nernst effects of magnons in altermagnets, Phys. Rev. B 108, L180401 (2023).

\bibitem{10} X. Zhou, W. Feng, R. W. Zhang, L. Smejkal, J. Sinova, Y. Mokrousov, and Y. Yao, Crystal Thermal Transport in Altermagnetic RuO$_2$, Phys. Rev. Lett. 132, 056701 (2024).

\bibitem{11} H. Reichlova, R. L. Seeger, R. G. Hernandez, I. Kounta, R. Schlitz, D. Kriegner, P. Ritzinger, M. Lammel, M. Leiviska, V. Petricek, P. Dolezal, E. Schmoranzerova, A. Badura, A. Thomas, V. Baltz, L. Michez, J. Sinova, S. T. B. Goennenwein, T. Jungwirth, and L. Smejkal, Macroscopic time reversal symmetry breaking by staggered spin-momentum interaction, arXiv:2012.15651.

\bibitem{12} L. Smejkal, A. B. Hellenes, R. G. Hernandez, J. Sinova, and T. Jungwirth, Giant and tunneling magnetoresistance in unconventional collinear antiferromagnets with nonrelativistic spin-momentum coupling, Phys. Rev. X 12, 011028 (2022).

\bibitem{13} X. Zhou, W. Feng, X. Yang, G. Y. Guo, and Y. Yao, Crystal chirality magneto-optical effects in collinear antiferromagnets, Phys. Rev. B 104, 024401 (2021).

\bibitem{14} Y. Fang, J. Cano, and S. A. A. Ghorashi, Quantum Geometry Induced Nonlinear Transport in Altermagnets, Phys. Rev. Lett. 133, 106701 (2024).

\bibitem{15} R. G. Hernandez, L. Smejkal, K. Vyborny, Y. Yahagi, J. Sinova, T. Jungwirth, and J. Zelezny, Efficient Electrical Spin Splitter Based on Nonrelativistic Collinear Antiferromagnetism, Phys. Rev. Lett. 126, 127701 (2021).

\bibitem{16} S. Karube, T. Tanaka, D. Sugawara, N. Kadoguchi, M. Kohda, and J. Nitta, Observation of spin-splitter torque in collinear antiferromagnetic RuO$_2$, Phys. Rev. Lett. 129, 137201 (2022).

\bibitem{17} H. Bai, L. Han, X. Y. Feng, Y. J. Zhou, R. X. Su, Q. Wang, L. Y. Liao, W. X. Zhu, X. Z. Chen, F. Pan, X. L. Fan, and C. Song, Observation of spin splitting torque in a collinear antiferromagnet RuO$_2$, Phys. Rev. Lett. 128, 197202 (2022).

\bibitem{18} S. A. A. Ghorashi, T. L. Hughes, and J. Cano, Altermagnetic Routes to Majorana Modes in Zero Net Magnetization, Phys. Rev. Lett. 133, 106601 (2024).

\bibitem{19} H. G. Giil, and J. Linder, Superconductor-altermagnet memory functionality without stray fields, Phys. Rev. B 109, 134511 (2024).

\bibitem{20} S. Lee, S. Lee, S. Jung, J. Jung, D. Kim, Y. Lee, B. Seok, J. Kim, B. G. Park, L. Smejkal, C. J. Kang, and C. Kim, Broken Kramers Degeneracy in Altermagnetic MnTe, Phys. Rev. Lett. 132, 036702 (2024).

\bibitem{21} Z. Liu, M. Ozeki, S. Asai, S. Itoh, and T. Masuda, Chiral split magnon in altermagnetic MnTe, Phys. Rev. Lett. 133, 156702 (2024).

\bibitem{22} J. Ding, Z. Jiang, X. Chen, Z. Tao, Z. Liu, T. Li, J. Liu, J. Sun, J. Cheng, J. Liu, Y. Yang, R. Zhang, L. Deng, W. Jing, Y. Huang, Y. Shi, M. Ye, S. Qiao, Y. Wang, Y. Guo, D. Feng, and D. Shen, Large Band Splitting in $\emph{g}$-Wave Altermagnet CrSb, Phys. Rev. Lett. 133, 206401 (2024).

\bibitem{23} M. Zeng, M. Y. Zhu, Y. P. Zhu, X. R. Liu, X. M. Ma, Y. J. Hao, P. Liu, G. Qu, Y. Yang, Z. Jiang, K. Yamagami, M. Arita, X. Zhang, T. H. Shao, Y. Dai, K. Shimada, Z. Liu, M. Ye, Y. Huang, Q. Liu, and C. Liu, Observation of Spin Splitting in Room-Temperature MetallicAntiferromagnet CrSb, Adv. Sci. 11, 2406529 (2024).

\bibitem{24} G. Yang, Z. Li, S. Yang, J. Li, H. Zheng, W. Zhu, Z. Pan, Y. Xu, S. Cao, W. Zhao, A. Jana, J. Zhang, M. Ye, Y. Song, L. H. Hu, L. Yang, J. Fujii, I. Vobornik, M. Shi, H. Yuan, Y. Zhang, Y. Xu, and Y. Liu, Three-dimensional mapping of the altermagnetic spin splitting in CrSb, Nat. Commun. 16, 1442 (2025).

\bibitem{25} I. I. Mazin, K. Koepernik, M. D. Johannes, R. G. Hernandez, and L. Smejkal, Prediction of unconventional magnetism in doped FeSb$_2$, Proc. Natl. Acad. Sci. 118, e2108924118 (2021).

\bibitem{26} Q. Tian, S. I. Vishkayi, M. B. Tagani, L. Zhang, Y. Tian, L. J. Yin, L. Zhang, and Z. Qin, Two-Dimensional Artificial Ge Superlattice Confining in Electronic Kagome Lattice Potential Valleys, Nano Lett. 23, 9851 (2023).

\bibitem{27} Q. Tian, M. B. Tagani, S. I. Vishkayi, C. Zhang, B. Li, L. Zhang, L. J. Yin, Y. Tian, L. Zhang, and Z. Qin, Twist-Angle Tuning of Electronic Structure in Two-Dimensional Dirac Nodal Line Semimetal Au$_2$Ge on Au(111), ACS Nano 28, 9011 (2024).

\bibitem{28} D. Xiao, W. Yao, and Q. Niu, Valley-Contrasting Physics in Graphene: Magnetic Moment and Topological Transport, Phys. Rev. Lett. 99, 236809 (2007).

\bibitem{29} T. Cao, G. Wang, W. Han, H. Ye, C. Zhu, J. Shi, Q. Niu, P. Tan, E. Wang, B. Liu, and J. Feng, Valley-selective circular dichroism of monolayer molybdenum disulphide, Nat. Commun. 3, 887 (2012).

\bibitem{30} P. Li, B. Liu, S. Chen, W. X. Zhang, and Z. X. Guo, Progress on two-dimensional ferrovalley materials, Chin. Phys. B 33, 017505 (2024).

\bibitem{31} W. Y. Tong, S. J. Gong, X. Wan, and C. G. Duan, Concepts of ferrovalley material and anomalous valley Hall effect, Nat. Commun. 7, 13612 (2016).

\bibitem{32} K. Wang, Y. Li, H. Mei, P. Li, and Z. X. Guo, Quantum anomalous Hall and valley quantum anomalous Hall effects in two-dimensional d$^0$ orbital XY monolayers, Phys. Rev. Mater. 6, 044202 (2022).

\bibitem{33} P. Li, C. Wu, C. Peng, M. Yang, and W. Xun, Multifield tunable valley splitting in two-dimensional MXene Cr$_2$COOH, Phys. Rev. B 108, 195424 (2023).

\bibitem{34} P. Li, X. Yang, Q. S. Jiang, Y. Z. Wu, and W. Xun, Built-in electric field and strain tunable valley-related multiple topological phase transitions in VSiXN$_4$ (X=C, Si, Ge, Sn, Pb) monolayers, Phys. Rev. Mater. 7, 064002 (2023).

\bibitem{35} S. D. Guo, L. Zhang, Y. Zhang, P. Li, and G. Wang, Large spontaneous valley polarization and anomalous valley Hall effect in antiferromagnetic monolayer Fe$_2$CF$_2$, Phys. Rev. B 110, 024416 (2024).

\bibitem{36} W. Xun, C. Wu, H. Sun, W. Zhang, Y. Z. Wu, and P. Li, Coexisting Magnetism, Ferroelectric, and Ferrovalley Multiferroic in Stacking-Dependent Two-Dimensional Materials, Nano Lett. 24, 3541 (2024).

\bibitem{37} G. Kresse, and J. Hafner. $\emph{Ab initio}$ molecular dynamics for liquid metals Phys. Rev. B 47, 558 (1993).

\bibitem{38} G. Kresse, and J. Furthmuller. Efficient iterative schemes for ab initio total-energy calculations using a plane-wave basis set, Phys. Rev. B 54, 11169 (1996).

\bibitem{39} G. Kresse, and D. Joubert. From ultrasoft pseudopotentials to the projector augmented-wave method, Phys. Rev. B 59, 1758 (1999).

\bibitem{40} J. P. Perdew, K. Burke, and M. Ernzerhof. Generalized Gradient Approximation Made Simple, Phys. Rev. Lett. 77, 3865 (1996).

\bibitem{41} S. Grimme, J. Antony, S. Ehrlich, and H. Krieg. A consistent and accurate ab initio parametrization of density functional dispersion correction (DFT-D) for the 94 elements H-Pu, J. Chem. Phys. 132, 154104 (2010).

\bibitem{42} W. Xun, X. Liu, Y. Zhang, Y. Z. Wu, and P. Li, Stacking-, Strain-Engineering Induced Altermagnetism, Multipiezo Effect, and Topological State in Two-Dimensional Materials, Appl. Phys. Lett. 126, 161903 (2025).

\bibitem{43} B. Goodenough, Theory of the Role of Covalence in the Perovskite-Type Manganites [La,(II)]MnO$_3$, Phys. Rev. 100, 564 (1955).

\bibitem{44} J. Kanamori, Superexchange interaction and symmetry properties of electron orbitals, J. Phys. Chem. Solids 10, 87 (1959).

\bibitem{45} P. W. Anderson, New Approach to the Theory of Superexchange Interactions, Phys. Rev. 115, 2 (1959).

\bibitem{46} P. Li, X. Li, J. Feng, J. Ni, Z. X. Guo, and H. Xiang, Origin of zigzag antiferromagnetic orders in XPS$_3$ (X = Fe, Ni) monolayers. Phys. Rev. B 109, 214418 (2024).

\bibitem{47} Y. Wang, H. Sun, C. Wu, W. Zhang, S. D. Guo, Y. She, and P. Li, Multifield tunable valley splitting and anomalous valley Hall effect in two-dimensional antiferromagnetic MnBr, Phys. Rev. B 111, 085432 (2025).

\bibitem{48} C. Wu, H. Sun, P. Dong, Y. Z. Wu, and P. Li, Coexisting Triferroic and Multiple Types of Valley Polarization by Structural Phase Transition in 2D Materials, Adv. Funct. Mater. 35, 2501506 (2025).

\bibitem{49} H. Sun, Y. Ren, C. Wu, P. Dong, W. Zhang, Y. Z. Wu, and P. Li, Ferroelectric tuning of the valley polarized metal-semiconductor transition in Mn$_2$P$_2$S$_3$Se$_3$/Sc$_2$CO$_2$ van der Waals heterostructures and application to nonlinear Hall effect devices, Phys. Rev. Appl. 23, 034032 (2025).

\bibitem{50} Y. Zhu, T. Chen, Y. Li, L. Qiao, X. Ma, C. Liu, T. Hu, H. Gao, and W. Ren, Multipiezo Effect in Altermagnetic V$_2$SeTeO Monolayer, Nano Lett. 24, 472 (2024).

\bibitem{51} Y. Jiang, X. Zhang, H. Bai, Y. Tian, B. Zhang, W. J. Gong, and X. Kong, Strain-engineering spin-valley locking effect in altermagnetic monolayer with multipiezo properties, Appl. Phys. Lett. 126, 053102 (2025).





	
\end{thebibliography}

\end{document}